\documentclass[conference]{IEEEtran}
\usepackage{blindtext, graphicx}
\usepackage[utf8]{inputenc}
\usepackage[english]{babel}
\usepackage{caption}
\usepackage{graphicx}
\usepackage{subcaption}
\usepackage{float}

\ifCLASSINFOpdf
\else
\fi
\begin{document}
%
\title{Data clustering using stochastic block models}

\author{\IEEEauthorblockN{Nina Mrzelj}
\IEEEauthorblockA{University of Ljubljana, \\
Faculty of Mathematics and Physics\\
Email: nina.mrzelj@gmail.com}
\and

\IEEEauthorblockN{Pavlin Gregor Poličar}
\IEEEauthorblockA{ University of Ljubljana,\\
Faculty of Computer and Information Science\\
Email: pavlin.g.p@gmail.com}}


%


\maketitle

\begin{abstract}
It has been shown that community detection algorithms work better for clustering tasks than other, more popular methods, such as k-means. In fact, network analysis based methods often outperform more widely used methods and do not suffer from some of the drawbacks we notice elsewhere e.g. the number of clusters $k$ usually has to be known in advance. However, stochastic block models which are known to perform well for community detection, have not yet been tested for this task. We discuss why these models cannot be directly applied to this problem and test the performance of a generalization of stochastic block models which work on weighted graphs and compare them to other clustering techniques.
\end{abstract}


%
\IEEEpeerreviewmaketitle

\section{Introduction}
Clustering methods are one of the fundamental topics in machine learning since they can be used to detect structure in unlabelled data. As this is such a broadly applicable task, many solutions have been proposed~\cite{theodoridis2003pattern}, the most popular of which is k-means~\cite{macqueen1967some}, due to its simplicity, fast execution time and good performance. However, k-means and other traditional machine learning approaches to clustering suffer from certain drawbacks. In many cases, the number of clusters $k$ must be predefined i.e. we must know in advance how many clusters appear in the data. Furthermore, most of these techniques produce clusters of circular geometric shapes (this is especially true for k-means). These are both undesirable properties, since we can rarely know the correct number of clusters in advance, or that clusters appear in a particular shape.

The idea to use community detection algorithms for the clustering task was first proposed in 2008~\cite{de2008data}. The suggested approach was robust in finding clusters of different shapes, but the number of clusters still had to be provided. However, the authors considered only a single community detection algorithm with a single metric.

In the work of Rodrigues et al.~\cite{rodrigues2011complex}, other community detection algorithms were considered with a variety of metrics. It was shown that certain community detection algorithms almost always outperform more traditional methods and that some of these approaches do not suffer from the aforementioned drawbacks of other methods. In particular, they showed that using Manhattan or Chebyshev dissimilarity is most suitable for complex network approaches for clustering. The smallest misclassification rates were achieved with a community detection method based on greedy modularity optimization.

We extend the research done by Rodrigues et al. with stochastic block models \cite{holland1983stochastic}, which are also known to work well for community detection tasks in the scope of network analysis~\cite{fortunato2010community}. We discuss the difficulties that arise when trying to apply these models to such data and provide a solution. We also solve the clustering task using a generalization of stochastic block models that support weighted graphs~\cite{aicher2013adapting}. We evaluate the performance of these models with a series of classification tasks, both on synthetic and real-world datasets and compare them to other complex network based methods, as well as more traditional clustering techniques.

\section{Concepts and methods}

Stochastic block models are known to work well for community detection and can reveal many forms of network structure, so naturally, we expect it to also work well for data clustering. However, they cannot be directly applied to clustering problems, since they cannot account for edge weights, which are the central element of clustering using a complex networks approach.

\subsection{Stochastic block models}

The stochastic block model (SBM)~\cite{holland1983stochastic} is a generative model for learning community structure in unweighted networks. The model assigns each node $i$ to one of $K$ blocks denoted by $z_i$, and each each edge $A_{ij}$ exists with a probability $\theta_{z_i z_j}$ where $\theta$ is a $K \times K$ matrix which values are the probabilities of links appearing between blocks $i$ and $j$. The optimal block assignment is selected using a probabilistic maximum likelihood approach.

The stochastic block model can generate a wide variety of patterns that appear in network analysis including assortative, disassortative and core-periphery structures. However, this probabilistic model assumes an unweighted network. To remedy this, weighted networks are most often simplified into unweighted networks by applying a single global threshold.

This is problematic, since Aicher et al.~\cite{aicher2013adapting} showed that there are cases where this approach will never yield the correct block structure for any thershold value. Applying a single global threshold completely discards any local information that may be encoded within the edge weights. They propose a generalization of the stochastic block model, the weighted stochastic block model (WSBM) which removes the need for thresholding altogether. The model assumes that the weights are exponentially distributed and accepts a parameter $\alpha$ which balances the information gained from a missing edge and an edge with zero weight.

\subsection{Representing tabular data as a network}

The first step of any clustering using complex networks approaches requires the tabular data to be represented as a graph. A full graph is induced from the data. Each vertex represents a data point and their distances to other nodes are represented as the weights on the corresponding edges. Since the space of possible metrics is vast, we limit ourselves to the following metrics:

\begin{enumerate}
    \item Exponential of Chebyshev distance
    $$ S_C(x,y) = {\textnormal{exp}( - max_{i = 1}^n \left | x_i - y_i \right | )} $$

    \item Exponential of Manhattan distance
    $$ S_M(x,y) = \textnormal{exp} \Big ( -\sum_{i=1}^{n} \left | x_i - y_i \right | \Big)$$
    
    \item Exponential of Euclidean distance
    $$ S_E(x,y) = \textnormal{exp} \Big ( - \big ( \sum_{i=1}^{n} \left ( x_i - y_i \right ) ^2 \big ) ^{\frac{1}{2}} \Big) $$

\end{enumerate}

In the formulae above, $x, y \in {\rm I\!R}^n$ denote data points with $n$ features. Metrics 1 and 2 are chosen because were shown to produce small error rates in work of Rodrigues et al.~\cite{rodrigues2011complex}. In networks, where these two measures work poorly, we also consider the exponential of Euclidean distance.

We only consider exponential distances since they have the desirable property of being limited to the interval $\left [ 0, 1 \right ]$. Moreover, a larger exponential distance indicates greater similarity, whereas the opposite is true for their non-exponential versions. This is conceptually more sound, since we would intuitively assign larger values to data points which are more similar.

Using only the exponential variants of the distance metrics, we also satisfy the requirement imposed by the WSBM: that the edge weights in a network must be exponentially distributed.

We can now directly use the induced graphs with algorithms that can take edge weights into account, such as the WSBM. However, the classic SBM cannot do this, therefore a preprocessing step is required to remove a number of edges, so that the model may find any meaningful structure within the data. Were we to ignore this step, the model would find a single block containing all the nodes. The simplest solution is to apply a single global threshold to all the edges, and to apply the model to the pruned graph. We describe this approach in the following section.

\subsection{Determine the optimal cutoff threshold}

\begin{figure}[H]
    \centering
    \begin{subfigure}[c]{\columnwidth}
        \includegraphics[width=\columnwidth]{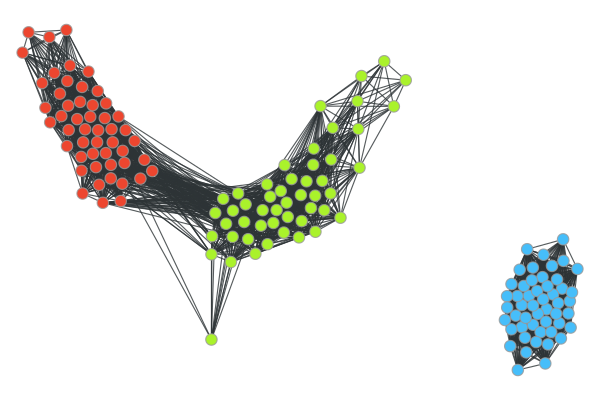}
        \caption{Threshold of 0.1}
        \label{fig:iris_graph1}
    \end{subfigure}
    ~
    \begin{subfigure}[c]{\columnwidth}
        \includegraphics[width=\columnwidth]{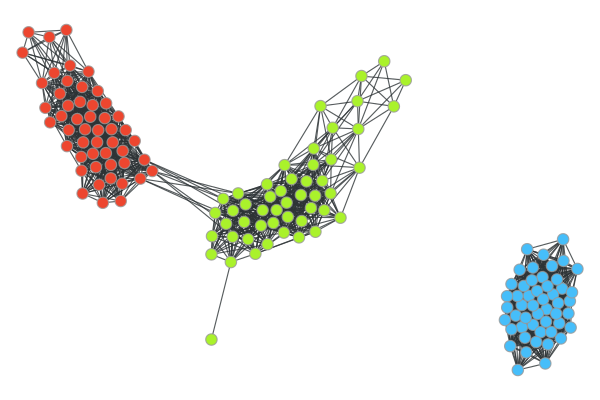}
        \caption{Threshold of 0.2}
        \label{fig:iris_graph2}
    \end{subfigure}
    ~
    \begin{subfigure}[c]{\columnwidth}
        \includegraphics[width=\columnwidth]{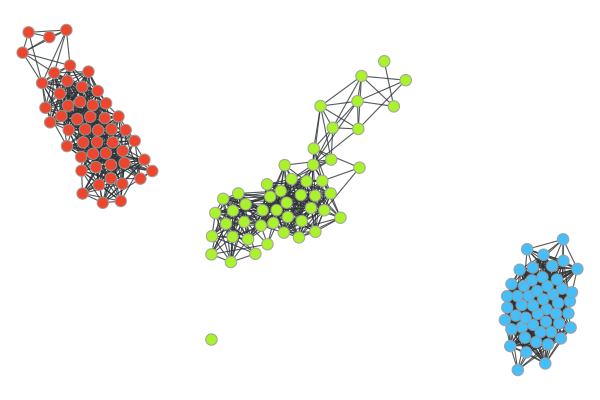}
        \caption{Threshold of 0.3}
        \label{fig:iris_graph3}
    \end{subfigure}
    \caption{The induced graph of the Iris dataset with a three different global thresholds. The dataset contains three distinct classes, which are clearly separated in the above images.}\label{fig:iris_graph}
\end{figure}

Before we can apply the classic SBM~\cite{karrer2011stochastic} on the graph, we must first find an appropriate threshold, so that the model may find any meaningful structure. However, finding such a threshold so that optimal clustering is achieved is not trivial. Figure~\ref{fig:iris_graph} shows the induced graph on the well-known Iris dataset for three different threshold values. We expect that the SBM may be able to detect the correct block structure for all three values of the cutoff rate.

This can be translated into the classical problem that other traditional clustering methods face, when the optimal number of clusters cannot be inferred automatically by the algorithm or is not known. We evaluate cluster labelling for different thresholds using the silhouette score (see Equation~\ref{eq:silhouette}) - a popular evaluation metric for evaluating clustering quality. 

\begin{equation}\label{eq:silhouette}
s(i) = \frac{{b(i) - a(i)}}{\max\{a(i),b(i)\}}
\end{equation}

The silhouette score is calculated with $a(i)$, which denotes the mean intra-cluster distance from point $i$ to all other points within that same cluster, and $b(i)$, which denotes the mean distance from point $i$ to all points within the nearest cluster to point $i$.

However, when the true cluster labels are known, we can use the more informative mutual information score or NMI (see Equation~\ref{eq:nmi}).

\begin{equation}\label{eq:nmi}
\textnormal{NMI}(T, C) = \frac{2 \, I(T, C)}{H(T) + H(C)}
\end{equation}

The NMI considers the mutual information $I$ of two clusterings: the true labels $T$ and our computed labels $C$. This is then normalized by the sum of their entropies $H$.  This score is a better indicator of the actual success rate of the method, since it also takes into account the true labels of the data.

\begin{figure}[ht]
    \centering
    \includegraphics[width=0.9\columnwidth]{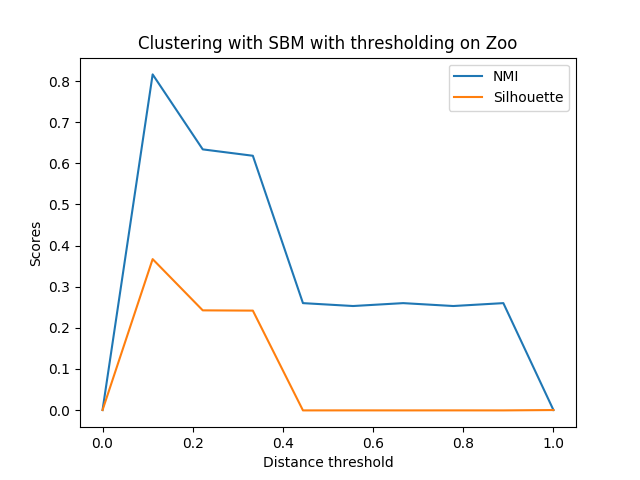}
    \caption{A comparison of silhouette scores and NMI scores on the Zoo dataset~\cite{Lichman:2013}.}
    \label{fig:threshold_sbm_iris_nmi_slh}
\end{figure}

By comparing the NMI and silhouette scores on the same datasets, we verify that indeed, the silhouette score is a good indicator of clustering quality (see Figure~\ref{fig:threshold_sbm_iris_nmi_slh}). They do not overlap, which is to be expected, as the scores have different bounds. However, it is clear that the optimal value is achieved at the same threshold value, therefore we can conclude that the silhouette score is a reliable indicator of cluster quality (this test was run on several datasets, but we include only one for illustrative purposes).

\section{Results and discussion}

\begin{figure}[H]
    \centering
    \begin{subfigure}[c]{0.3\columnwidth}
        \includegraphics[width=\columnwidth]{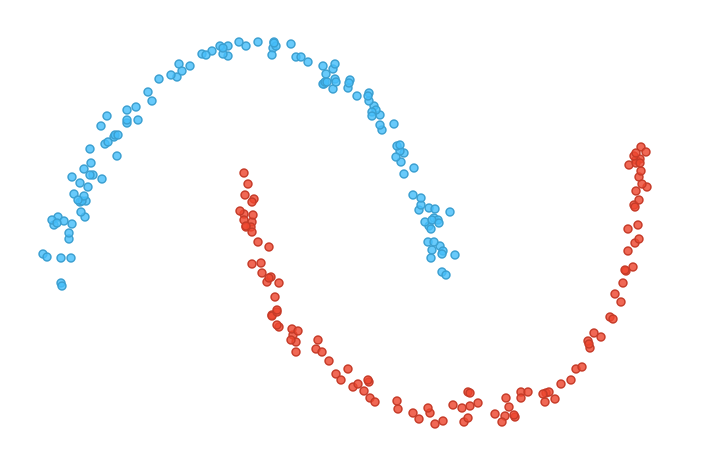}
        \caption{Two moons}
        \label{fig:synthetic_two_moons}
    \end{subfigure}
    ~
    \begin{subfigure}[c]{0.3\columnwidth}
        \includegraphics[width=\columnwidth]{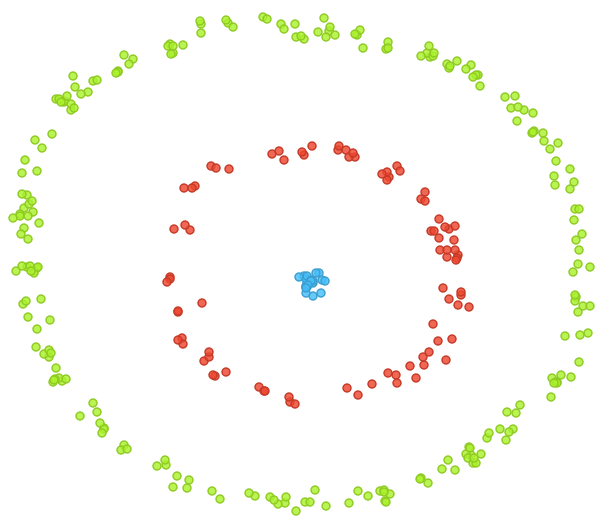}
        \caption{Circles}
        \label{fig:synthetic_circles}
    \end{subfigure}
    ~
    \begin{subfigure}[c]{0.3\columnwidth}
        \includegraphics[width=\columnwidth]{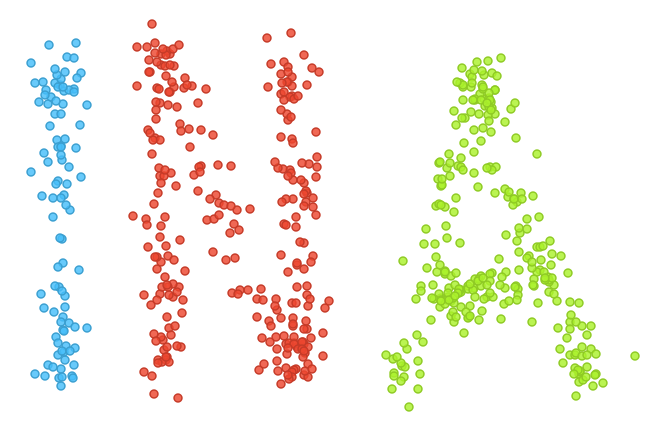}
        \caption{INA}
        \label{fig:synthetic_ina}
    \end{subfigure}
    \caption{Synthetic datasets}\label{fig:synthetic_datasets}
\end{figure}

We evaluate the proposed clustering techniques on five real-world datasets from the UCI Machine Learning Repository~\cite{Lichman:2013}: Iris, Ecoli, Glass, Zoo and Movements. We also test the methods on three synthetic datasets: two moons, circles and the INA dataset (shown in Figure~\ref{fig:synthetic_datasets}).

The dataset characteristics are summarized in Table~\ref{tab:datasets}. 

\begin{table}[ht]
    \centering
    \begin{tabular}{c c c c}
        Name & Instances & Features & Clusters\\ \hline
        iris & 150 & 4 & 3 \\
        ecoli & 336 & 7 & 8 \\
        glass & 214 & 9 & 6 \\
        zoo & 101 & 16 & 7 \\ 
        movements & 360 & 90 & 15 \\ \hline
        two moons & 250 & 2 & 2 \\
        circles & 336 & 2 & 3 \\
        ina & 660 & 2 & 3 \\
    \end{tabular}
    \caption{Description of datasets}
    \label{tab:datasets}
\end{table}

\subsection{Results}

We compare the SBM with thresholding and the WSBM with two popular clustering approaches: k-means and hierarchical clustering with Ward linkage. Since none of these methods can infer the optimal number of clusters automatically, we select the best clusterings with regard to the silhouette score.

To evaluate different clustering methods, we use three popular metrics: the adjusted Rand index or ARI~\cite{hubert1985comparing} (see Equation ~\ref{eq:ari}), normalized mutual information score (see Equation~\ref{eq:nmi}) and the silhouette score (see Equation~\ref{eq:silhouette}). The ARI is a variant of the Rand index that is corrected for chance and is used to measure the similarity between two clusterings. 

\begin{equation}\label{eq:ari}
\textnormal{ARI} = \frac{RI - ExpectedRI}{max(RI) - ExpectedRI}
\end{equation}

We present the most interesting results obtained in Table~\ref{tab:results} (the full results are not included here for the sake of brevity, but the interested reader can find them in Table~\ref{tab:results_full}). We include only the variants using the Manhattan distance metric, as it gave us the best results on average. We also include the results of the WSBM where the optimal number of clusters in known in advance, again using only the Manhattan distance metric as a measure of similarity.

\begin{table}[ht]
  \centering
  \begin{tabular}{c | c | c c c | c}
    Dataset & Method & Silhouette & NMI & ARI & Clusters \\ \hline
    Iris & k-means & \textbf{0.6808} & 0.6793 & 0.5399 & 2 \\
    Iris & Hierarchical clustering & \textbf{0.6864} & \textbf{0.7612} & \textbf{0.5681} & 2 \\
    Iris & SBM & 0.3916 & 0.6752 & \textbf{0.5703} & 7 \\
    Iris & WSBM & \textbf{0.6864} & \textbf{0.7612} & \textbf{0.5681} & 2 \\
    \textit{Iris} & \textit{WSBM (known)} & \textit{0.5171} & \textit{0.7206} & \textit{0.6906} & \textit{3} \\
    \hline
    Ecoli & k-means & \textbf{0.4303} & 0.6548 & 0.6860 & 4 \\
    Ecoli & Hierarchical clustering & 0.4166 & 0.6537 & 0.6575 & 4 \\
    Ecoli & SBM & 0.0694 & 0.4821 & 0.1801 & 20 \\
    Ecoli & WSBM & 0.4263 & \textbf{0.6827} & \textbf{0.7261} & 4 \\
    \textit{Ecoli} & \textit{WSBM (known)} & \textit{0.2022} & \textit{0.5724} & \textit{0.4363} & \textit{8} \\
    \hline
    Zoo & k-means & 0.4177 & 0.8172 & \textbf{0.8418} & 6 \\
    Zoo & Hierarchical clustering & \textbf{0.4288} & \textbf{0.8485} & \textbf{0.8483} & 7 \\
    Zoo & SBM & 0.3811 & \textbf{0.8503} & 0.8137 & 5 \\
    Zoo & WSBM & 0.3191 & 0.7939 & 0.5243 & 11 \\
    \textit{Zoo} & \textit{WSBM (known)} & \textit{0.1258} & \textit{0.6229} & \textit{0.3602} & \textit{7} \\
    \hline
    Two Moons & k-means & \textbf{0.5753} & \textbf{0.5795} & \textbf{0.2533} & 8 \\
    Two Moons & Hierarchical clustering & \textbf{0.5717} & \textbf{0.5807} & \textbf{0.2581} & 8 \\
    Two Moons & SBM & 0.5293 & 0.4944 & 0.1180 & 18 \\
    Two Moons & WSBM & \textbf{0.5692} & \textbf{0.5800} & \textbf{0.2550} & 8 \\
    \textit{Two Moons} & \textit{WSBM (known)} & \textit{0.4686} & \textit{0.2063} & \textit{0.2675} & \textit{2} \\
    \hline
    INA & k-means & \textbf{0.5466} & \textbf{0.8238} & \textbf{0.7538} & 2 \\
    INA & Hierarchical clustering  & \textbf{0.5466} & \textbf{0.8238} & \textbf{0.7538} & 2 \\
    INA & SBM & 0.4015 & 0.5468 & 0.1140 & 32 \\
    INA & WSBM & \textbf{0.5466} & \textbf{0.8238} & \textbf{0.7538} & 2 \\
    \textit{INA} & \textit{WSBM (known)} & \textit{0.4463} & \textit{0.7529} & \textit{0.6966} & \textit{3} \\
    \hline
    Circular & k-means & 0.4279 & 0.2706 & 0.0747 & 7 \\
    Circular & Hierarchical clustering & 0.4045 & 0.3324 & \textbf{0.0813} & 9 \\
    Circular & SBM & \textbf{0.5485} & \textbf{0.4901} & 0.0666 & 28 \\
    Circular & WSBM & 0.4178 & 0.3335 & \textbf{0.0894} & 8 \\
    \textit{Circular} & \textit{WSBM (known)} & \textit{0.2619} & \textit{0.2367} & \textit{0.1167} & \textit{3} \\
  \end{tabular}
  \caption{Results}
  \label{tab:results}
\end{table}

The results clearly indicate that our naive approach of applying the SBM with a simple thresholding scheme does not perform consistently. Sometimes, it outperforms the other methods by a large margin and other times, it does significantly worse. We also notice that this method most often produces a very large number of clusters.

On the other hand, the WSBM does reasonably well on most datasets, either achieving similar results to other state-of-the-art methods, while in other cases outperforming them. However, it does not show the promise of the large improvements in accuracy as presented by Rodrigues et al. The WSBM cannot fully detect the correct clusters even when the correct number of clusters is provided in advance, similarly to other techniques.

\begin{figure}[H]
    \centering
    \begin{subfigure}[c]{0.48\columnwidth}
        \includegraphics[width=\columnwidth]{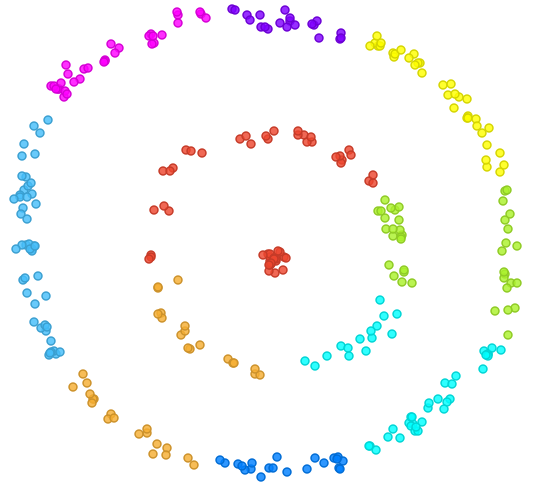}
        \caption{k-means}
        \label{fig:synthetic_two_moons}
    \end{subfigure}
    ~
    \begin{subfigure}[c]{0.48\columnwidth}
        \includegraphics[width=\columnwidth]{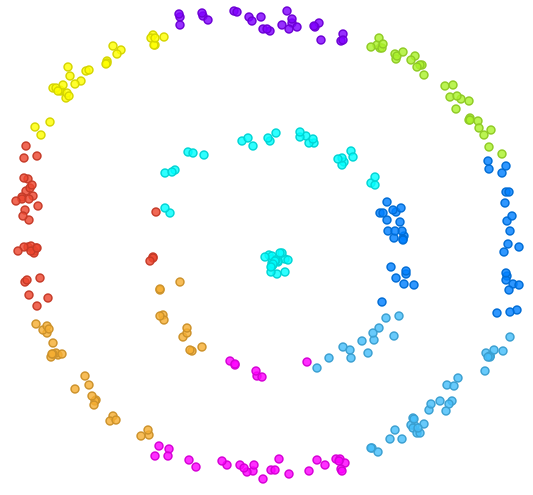}
        \caption{Hierarchical clustering}
        \label{fig:synthetic_circles}
    \end{subfigure}
    ~
    \begin{subfigure}[c]{0.48\columnwidth}
        \includegraphics[width=\columnwidth]{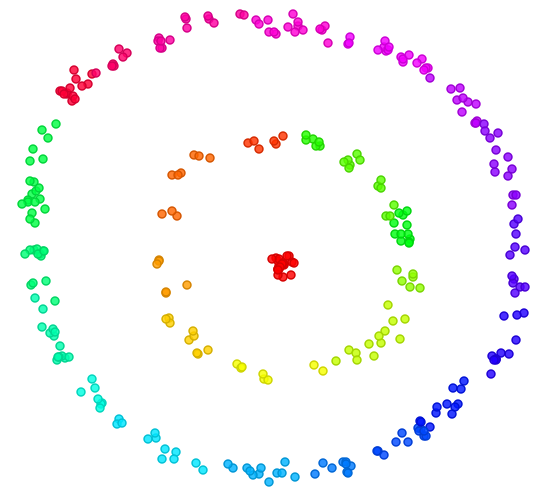}
        \caption{SBM with thresholding}
        \label{fig:synthetic_ina}
    \end{subfigure}
    ~
    \begin{subfigure}[c]{0.48\columnwidth}
        \includegraphics[width=\columnwidth]{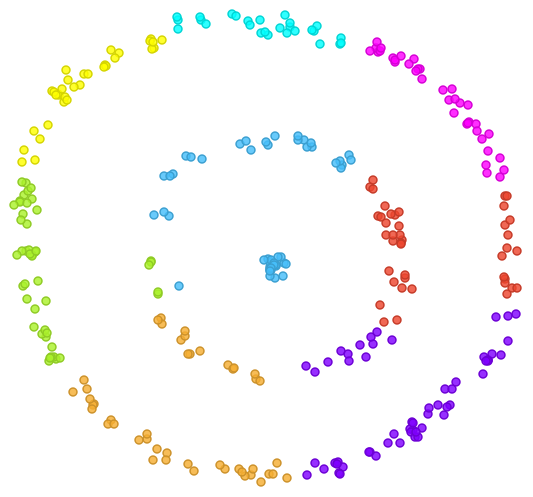}
        \caption{WSBM}
        \label{fig:synthetic_ina}
    \end{subfigure}
    \caption{Different clusterings for the Circles dataset.}\label{fig:results_circular}
\end{figure}

\begin{figure}[h]
    \centering
    \begin{subfigure}[c]{0.48\columnwidth}
        \includegraphics[width=\columnwidth]{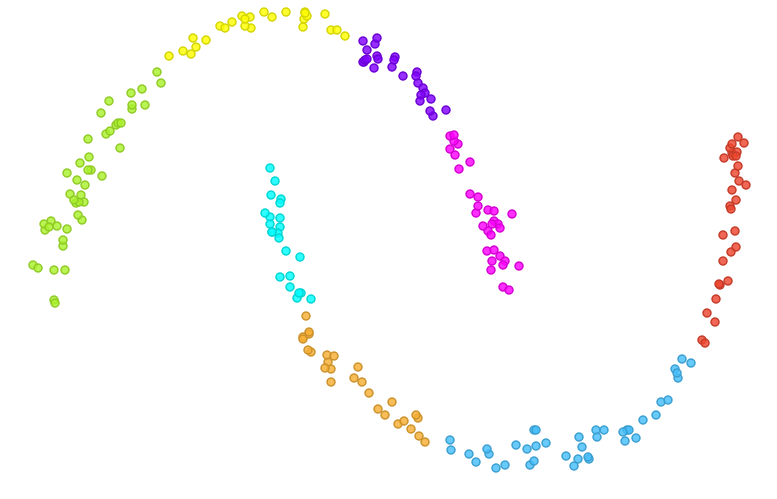}
        \caption{k-means}
        \label{fig:synthetic_two_moons}
    \end{subfigure}
    ~
    \begin{subfigure}[c]{0.48\columnwidth}
        \includegraphics[width=\columnwidth]{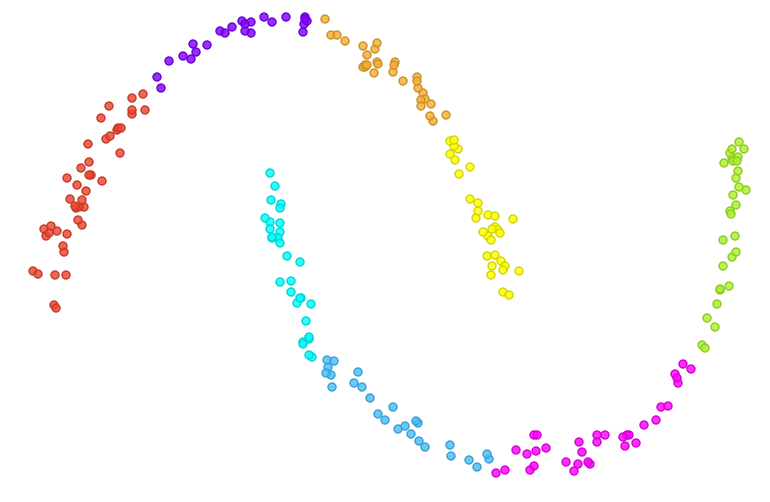}
        \caption{Hierarchical clustering}
        \label{fig:synthetic_circles}
    \end{subfigure}
    ~
    \begin{subfigure}[c]{0.48\columnwidth}
        \includegraphics[width=\columnwidth]{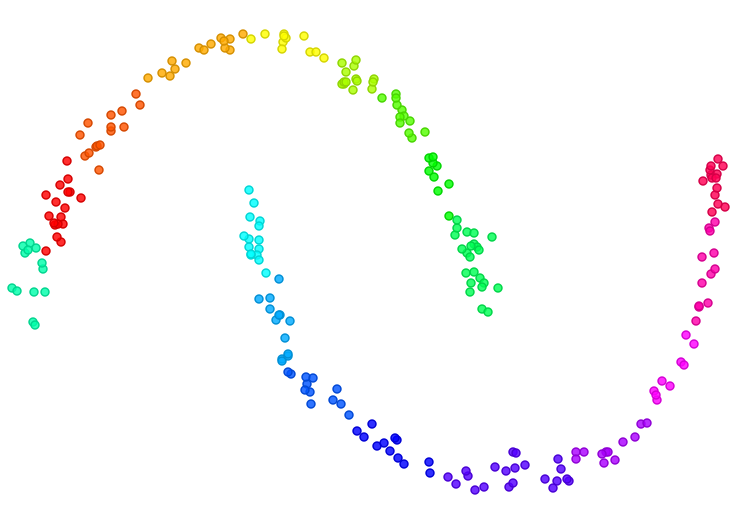}
        \caption{SBM with thresholding}
        \label{fig:synthetic_ina}
    \end{subfigure}
    ~
    \begin{subfigure}[c]{0.48\columnwidth}
        \includegraphics[width=\columnwidth]{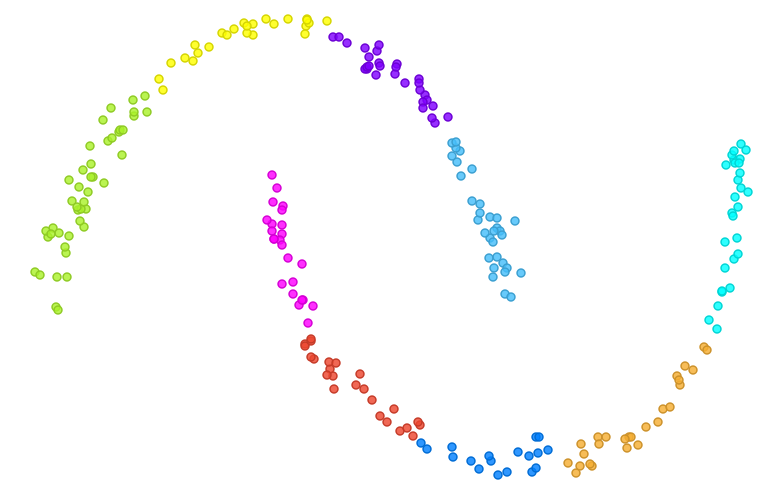}
        \caption{WSBM}
        \label{fig:synthetic_ina}
    \end{subfigure}
    \caption{Different clusterings for the Two Moons dataset.}\label{fig:results_moons}
\end{figure}

To attempt to visually understand how the approaches differ, we visualize the synthetic datasets (we do this for the synthetic datasets as they are inherently two dimensional and thus easily visualized). We show the labellings for the Circles datasets (see Figure~\ref{fig:results_circular}) and the Two Moons dataset (see Figure~\ref{fig:results_moons}). 

It is clear from the images that there are no apparent differences between hierarchical clustering, k-means and the WSBM. We again notice that the SBM has produced a very large number of clusters, however the clusters seem very similar to those produced by the other methods.

The poor results could in part be explained by the fact that we are trying to optimize the silhouette score. However, the silhouette score may not be the best indicator of our synthetic datasets e.g. circles, since the silhouette takes into account the mean distance to all nodes in the same cluster. We can imagine that this does not bode well for this particular dataset since the points that should be part of the same cluster can be very far apart (on opposite sides of the outer circle). Trying to maximize a different score e.g. the NMI might produce better results, but as already mentioned, we can only do this when we already know the true labels of the data, therefore this is not useful for unsupervised learning tasks.

Different evaluation metrics would produce different results and one may not be appropriate in every scenario. The best selection of the evaluation metric can rarely be known in advance. This is due to the fact that clustering is an ill-defined problem since the results of clustering is usually up for interpretation.

\section{Conclusion}

In the presented work, we study stochastic block models and their generalization for data clustering. We show that a naive approach with thresholding does not perform as consistently as other methods. On the other hand, weighted stochastic block models achieve state-of-the-art results. However, we do not observe a significant improvement in the quality of the discovered clusters. In addition, the proposed method does not overcome the problem of automatically inferring the correct number of clusters. We observe that different similarity metrics used in the graph induction step do not have a large impact on the results. The analysis of this work can be extended by comparing our results with other community detection methods and using other approaches for optimizing the cluster quality.

\bibliographystyle{ieeetr}
\bibliography{references}

\begin{thebibliography}{10}

\bibitem{theodoridis2003pattern}
S.~Theodoridis and K.~Koutroumbas, ``Pattern recognition,'' 2003.

\bibitem{macqueen1967some}
J.~MacQueen {\em et~al.}, ``Some methods for classification and analysis of
  multivariate observations,'' in {\em Proceedings of the fifth Berkeley
  symposium on mathematical statistics and probability}, vol.~1, pp.~281--297,
  Oakland, CA, USA., 1967.

\bibitem{de2008data}
T.~B. de~Oliveira, L.~Zhao, K.~Faceli, and A.~C. de~Carvalho, ``Data clustering
  based on complex network community detection,'' in {\em Evolutionary
  Computation, 2008. CEC 2008.(IEEE World Congress on Computational
  Intelligence). IEEE Congress on}, pp.~2121--2126, IEEE, 2008.

\bibitem{rodrigues2011complex}
F.~A. Rodrigues, G.~F. de~Arruda, and L.~d.~F. Costa, ``A complex networks
  approach for data clustering,'' {\em arXiv preprint arXiv:1101.5141}, 2011.

\bibitem{holland1983stochastic}
P.~W. Holland, K.~B. Laskey, and S.~Leinhardt, ``Stochastic blockmodels: First
  steps,'' {\em Social networks}, vol.~5, no.~2, pp.~109--137, 1983.

\bibitem{fortunato2010community}
S.~Fortunato, ``Community detection in graphs,'' {\em Physics reports},
  vol.~486, no.~3, pp.~75--174, 2010.

\bibitem{aicher2013adapting}
C.~Aicher, A.~Z. Jacobs, and A.~Clauset, ``Adapting the stochastic block model
  to edge-weighted networks,'' {\em arXiv preprint arXiv:1305.5782}, 2013.

\bibitem{karrer2011stochastic}
B.~Karrer and M.~E. Newman, ``Stochastic blockmodels and community structure in
  networks,'' {\em Physical Review E}, vol.~83, no.~1, p.~016107, 2011.

\bibitem{Lichman:2013}
M.~Lichman, ``{UCI} machine learning repository,'' 2013.

\bibitem{hubert1985comparing}
L.~Hubert and P.~Arabie, ``Comparing partitions,'' {\em Journal of
  classification}, vol.~2, no.~1, pp.~193--218, 1985.

\end{thebibliography}

\begin{table*}[t]
  \centering
  \begin{tabular}{c | c | c c c | c}
    Dataset & Method & Silhouette & NMI & ARI & clusters \\ \hline
    Iris & k-means & \textbf{0.6808} & 0.6793 & 0.5399 & 2 \\
    Iris & Hierarchical clustering & \textbf{0.6864} & \textbf{0.7612} & \textbf{0.5681} & 2 \\
    Iris & SBM (Manhattan) & 0.3916 & 0.6752 & 0.5703 & 7 \\
    Iris & SBM (Euclidean) & 0.3855 & 0.6283 & 0.5517 & 7 \\
    Iris & SBM (Chebyshev) & 0.3889 & 0.6661 & \textbf{0.5673} & 6 \\
    Iris & WSBM (Manhattan) & \textbf{0.6864} & \textbf{0.7612} & \textbf{0.5681} & 2 \\
    Iris & WSBM (Euclidean) & \textbf{0.6864} & \textbf{0.7612} & \textbf{0.5681} & 2 \\
    Iris & WSBM (Chebyshev) & \textbf{0.6864} & \textbf{0.7612} & \textbf{0.5681} & 2 \\

    \hline
    Ecoli & k-means & 0.4303 & 0.6548 & 0.6860 & 4 \\
    Ecoli & Hierarchical clustering & 0.4166 & 0.6537 & 0.6575 & 4 \\
    Ecoli & SBM (Manhattan) & 0.0694 & 0.4821 & 0.1801 & 20 \\
    Ecoli & SBM (Euclidean) & 0.1458 & 0.1964 & -0.0182 & 3 \\
    Ecoli & SBM (Chebyshev) & \textbf{0.5524} & 0.2057 & 0.0380 & 2 \\
    Ecoli & WSBM (Manhattan) & 0.4263 & \textbf{0.6827} & \textbf{0.7261} & 4 \\
    Ecoli & WSBM (Euclidean) & 0.4288 & \textbf{0.6860} & \textbf{0.7310} & 4 \\
    Ecoli & WSBM (Chebyshev) & 0.4115 & 0.6425 & 0.6803 & 4 \\

    \hline
    Zoo & k-means & 0.4177 & 0.8172 & \textbf{0.8418} & 6 \\
    Zoo & Hierarchical clustering & \textbf{0.4288} & \textbf{0.8485} & \textbf{0.8483} & 7 \\
    Zoo & SBM (Manhattan) & 0.3811 & \textbf{0.8503} & 0.8137 & 5 \\
    Zoo & SBM (Euclidean) & 0.3999 & 0.8011 & 0.8089 & 5 \\
    Zoo & SBM (Chebyshev) & 0.3065 & 0.4778 & 0.3143 & 3 \\
    Zoo & WSBM (Manhattan) & 0.3191 & 0.7939 & 0.5243 & 11 \\
    Zoo & WSBM (Euclidean) & 0.3229 & 0.7481 & 0.5077 & 10 \\
    Zoo & WSBM (Chebyshev) & 0.3317 & 0.6016 & 0.4964 & 4 \\
    
    \hline
    Glass & k-means & \textbf{0.5891} & \textbf{0.4409} & \textbf{0.2549} & 4 \\
    Glass & Hierarchical clustering & \textbf{0.5888} & 0.3953 & 0.2311 & 4 \\
    Glass & SBM (Manhattan) & 0.1187 & 0.3546 & 0.1308 & 9 \\
    Glass & SBM (Euclidean) & 0.1042 & 0.3662 & 0.1619 & 8 \\
    Glass & SBM (Chebyshev) & 0.3398 & 0.3615 & 0.2345 & 5 \\
    Glass & WSBM (Manhattan) & 0.3864 & 0.2303 & 0.1368 & 2 \\
    Glass & WSBM (Euclidean) & 0.4401 & 0.2811 & 0.1898 & 2 \\
    Glass & WSBM (Chebyshev) & 0.5446 & 0.3612 & 0.2259 & 2 \\
    
    \hline
    Movements & k-means & \textbf{0.2399} & 0.4750 & 0.2385 & 8 \\
    Movements & Hierarchical clustering & \textbf{0.2400} & 0.5219 & 0.2241 & 9 \\
    Movements & SBM (Manhattan) & -0.0842 & 0.1402 & 0.0052 & 2 \\
    Movements & SBM (Euclidean) & 0.1497 & 0.6213 & \textbf{0.3114} & 24 \\
    Movements & SBM (Chebyshev) & 0.1094 & \textbf{0.6357} & 0.2575 & 37 \\
    Movements & WSBM (Manhattan) & -0.0148 & 0.0923 & 0.0074 & 2 \\
    Movements & WSBM (Euclidean) & 0.1975 & 0.2579 & 0.0639 & 2 \\
    Movements & WSBM (Chebyshev) & 0.0846 & 0.4798 & 0.2517 & 7 \\

    \hline
    Two Moons & k-means & \textbf{0.5753} & 0.5795 & 0.2533 & 8 \\
    Two Moons & Hierarchical clustering & \textbf{0.5717} & 0.5807 & 0.2581 & 8 \\
    Two Moons & SBM (Manhattan) & 0.5293 & 0.4944 & 0.1180 & 18 \\
    Two Moons & SBM (Euclidean) & 0.5489 & 0.5162 & 0.1486 & 14 \\
    Two Moons & SBM (Chebyshev) & 0.4732 & 0.4937 & 0.1198 & 19 \\
    Two Moons & WSBM (Manhattan) & 0.5692 & 0.5800 & 0.2550 & 8 \\
    Two Moons & WSBM (Euclidean) & 0.5656 & 0.5654 & 0.2316 & 9 \\
    Two Moons & WSBM (Chebyshev) & 0.5377 & \textbf{0.6227} & \textbf{0.3322} & 6 \\
    \hline
    INA & k-means & \textbf{0.5466} & \textbf{0.8238} & \textbf{0.7538} & 2 \\
    INA & Hierarchical clustering  & \textbf{0.5466} & \textbf{0.8238} & \textbf{0.7538} & 2 \\
    INA & SBM (Manhattan) & 0.4015 & 0.5468 & 0.1140 & 32 \\
    INA & SBM (Euclidean) & 0.3780 & 0.5458 & 0.1128 & 33 \\
    INA & SBM (Chebyshev) & 0.2946 & 0.5320 & 0.1132 & 35 \\
    INA & WSBM (Manhattan)  & \textbf{0.5466} & \textbf{0.8238} & \textbf{0.7538} & 2 \\
    INA & WSBM (Euclidean)  & \textbf{0.5466} & \textbf{0.8238} & \textbf{0.7538} & 2 \\
    INA & WSBM (Chebyshev)  & \textbf{0.5466} & \textbf{0.8238} & \textbf{0.7538} & 2 \\
    
    \hline
    Circular & k-means & 0.4279 & 0.2706 & 0.0747 & 7 \\
    Circular & Hierarchical clustering & 0.4045 & 0.3324 & 0.0813 & 9 \\
    Circular & SBM (Manhattan) & 0.5485 & \textbf{0.4901} & 0.0666 & 28 \\
    Circular & SBM (Euclidean) & 0.5257 & \textbf{0.4941} & 0.0704 & 27 \\
    Circular & SBM (Chebyshev) & \textbf{0.5635} & \textbf{0.4977} & 0.0751 & 26 \\
    Circular & WSBM (Manhattan) & 0.4178 & 0.3335 & 0.0894 & 8 \\
    Circular & WSBM (Euclidean) & 0.4264 & 0.3281 & 0.0846 & 8 \\
    Circular & WSBM (Chebyshev) & 0.4124 & 0.3261 & \textbf{0.1143} & 6 \\

  \end{tabular}
  \caption{Results}
  \label{tab:results_full}
\end{table*}

\begin{IEEEbiography}[{\includegraphics[width=1in,height=1.25in,clip,keepaspectratio]{picture}}]{John Doe}
\blindtext
\end{IEEEbiography}

\end{document}